\documentclass[12pt]{article}
\usepackage[latin1]{inputenc}
\usepackage{graphicx,epsfig}
\usepackage{amsmath}
\usepackage{amsfonts}
\usepackage{amssymb}
\usepackage{amsthm}
\usepackage{bm}
\def\be{\begin{equation}}
\def\ee{\end{equation}}
\def\ba{\begin{eqnarray}}
\def\ea{\end{eqnarray}}

\begin{document}

\thispagestyle{empty}
\begin{center}

\vspace{0.4cm}
{\Large {\bf An analog fluid model for some tachyonic effects in field theory }}\\
\vspace{1.0cm}
{\large {\bf E. Arias\footnote{e-mail: enrike@cbpf.br}}, {\bf C. H. G. B\'essa\footnote{e-mail: chgbessa@cbpf.br} }, {\bf N. F. Svaiter\footnote{e-mail: nfuxsvai@cbpf.br} }} \\
\vspace{0.2cm}
{\it Centro Brasileiro de Pesquisas F\'{\i}sicas, Rua Dr. Xavier Sigaud 150, 22290-180, Rio de Janeiro, RJ, Brazil.} \\

\vspace{1.0cm}
\end{center}

\begin{abstract}
\begin{description}

We consider the sound radiation from an acoustic point-like source moving along a supersonic (``space-like") 
trajectory in a fluid at rest. We call it an acoustic ``tachyonic" source. We describe the radiation 
emitted by this supersonic source. After quantizing the acoustic perturbations, we present the 
distribution of phonons generated by this classical tachyonic source and the classical wave interference pattern.

\end{description}
\end{abstract}



\section{INTRODUCTION}


The concept of space-time is based in the geometric construction of fixed light-cones, which divide space-time into causally distinct regions.
Acoustic perturbations in a fluid define discontinuity surfaces that provide, for observers at rest with respect to the fluid, a causal structure with sound cones. Using the fact that fluctuations of the metric can lead to fluctuations in the light-cone \cite{f95}, \cite{fs96}, \cite{fs97} and \cite{ysf09}, it was proposed in condensed matter physics an analog model for quantum gravity effects \cite{kms10}. Further, a scalar quantum field theory in disordered media was investigated \cite{agkms11}. In turn, Unruh has shown that the propagation of sound waves in a supersonic fluid is equivalent to the propagation of scalar waves in a black hole space-time (\cite{u81}, see also Ref. \cite{su07}). On the other hand, in the last forty years several works treating different aspects of tachyons (super-luminal particles) were considered in the literature. Some of them deal with the development of a classical theory of tachyons and also with some  possible phenomenological applications \cite{r74, rm74, mr74, rrr86}: see also \cite{r09} and references therein. Astrophysical aspects, e.g.,  were considered in \cite{natureb77},\cite{natures}, \cite{naturep81}, \cite{gmmr86} and \cite{rcmr86}; while more recently one of the major interest in real tachyonic fields refers to the dark energy theory, where a tachyonic field is one possible candidate for this phenomenon \cite{s02}, \cite{scq04}, \cite{s07}, \cite{d08}, \cite{spq09} and \cite{cd11}. Other aspects are related with the study of Nondiffracting Waves, called Localized Waves \cite{rrb10, rrd04}, and of superluminal tunneling \cite{aer02} and \cite{r04}. In the first case, the experimental production of the so-called X-shaped waves was first performed by Lu et al. just in the acoustic case, when the X-waves are of course supersonics rather than superluminal \cite{lg92}. The first production of electromagnetic superluminal X-waves was done by Saari in Optics \cite{sr97}. From the theoretical point of view, among the first papers to predict X-waves let us quote \cite{bmr82} and \cite{r98}.  While, in the tunneling case, Recami et al. found that the total traversal time through quantum or classical barriers does not depend on the barrier width [``Hartman Effect", implying superluminal tunneling for long enough (opaque) barriers]; nor, in the case of two or more successive barriers, on the separation between them \cite{or92}, \cite{orrz95}, \cite{ors02} and \cite{orj04}. The experimental confirmation of the latter ``Generalized Hartman Effect" this effect can be found in \cite{nes94} and \cite{llbr02}. Indeed, other relevant experiments can be seen in \cite{r08, r09}. 
However the quantum field theory for tachyons is still not well understood, despite of some seminal papers  \cite{f67}, \cite{ds68} and \cite{tss81}, where the authors treated some quantum field theory aspects of tachyons in flat and curved space-time.

In the present paper, we study an analog model for a tachyonic source using fluid dynamics. Some quantum field theory aspects of the model are discussed. More precisely, we consider a fluid at rest and the sound radiation from a classical acoustic point-like source moving along a space-like trajectory (supersonic trajectory). After quantizing the acoustic perturbations, the distribution of phonons generated by this classical supersonic source is presented. This picture mimics one well known phenomenon in tachyon physics, that, is the double image effect which can be seen by a sub-luminal observer in the presence of a super-luminal source, one set of the images being received in the reversed chronological order \cite{r74}. The same situation does indeed occur also for a supersonic source motion in a fluid at rest, where now the static observer listen a double sound emitted by the supersonic source. Other effects are found, like the confinement of the acoustic radiation inside a double cone, which also mimics what
known to happen in the analogous superluminal cases \cite{rcmr86, rrd04, r98}. 

The organization of the paper is as follows. In Section \ref{sec2} we discuss the sound radiation from an acoustic source moving supersonically, i.e. along a space-like trajectory . In Section 3, the distribution of phonons generated by this classical supersonic source is presented. At the end of this section we also discus about the interference pattern measured by the observer at rest at a fixed instant of time. This last effect occurs only 
in the supersonic case. Our results are summarized and discussed in the conclusions. In this paper we use $\hbar =  k_B = 1$

\section{THE SOUND RADIATION FROM AN ACOUSTIC SOURCE}\label{sec2}

Aim of this section is to describe the sound radiation emitted by an acoustic source that moves along a trajectory with velocity greater than the sound speed, in a fluid at rest. To this purpose we follow basically Ref. \cite{mi68}.

It is well known that the flow field about a source of finite dimensions is turbulent. To avoid this effect, let us consider the idealized situation of an acoustic point source. This simplification also avoids the question of high frequencies, that arises when the wavelength of the field is of the order of the source dimensions.

We shall first study the kinematical properties of the system. Let us consider the sound radiation from an acoustic point that moves along a generic trajectory in a fluid at rest. The path of the source is specified by the vector: ${\bf{r}}_s(t) = (x_s(t), y_s(t), z_s(t))$. The point of observation ``O" has the coordinate ${\bf{r}}$ given by: ${\bf{r}} = (x, y, z)$.

\begin{figure}
\begin{center}
\includegraphics[width=0.4\textwidth]{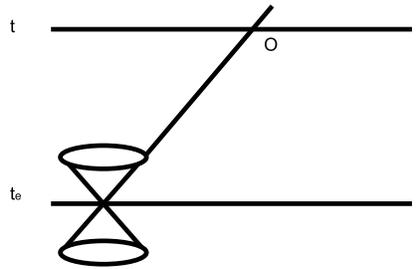}
\end{center}
\caption{Source in uniform subsonic motion. }\label{fig1}
\end{figure}

The sound pressure observed at $\vec{r}$ at the time $t$, emitted by the source at the time $t_e = t - R/c$, when the source was at the emission point $E$, that is, at $\vec{r}_s(t_e)$. \ We have ${\bf{r}} = {\bf{R}} + {\bf{r}}_s$, and therefore ${\bf{R}} = {\bf{r}} - {\bf{r}}_s(t_e)$. The distance $R$ can be written as

\begin{figure}
\begin{center}
\includegraphics[width=0.4\textwidth]{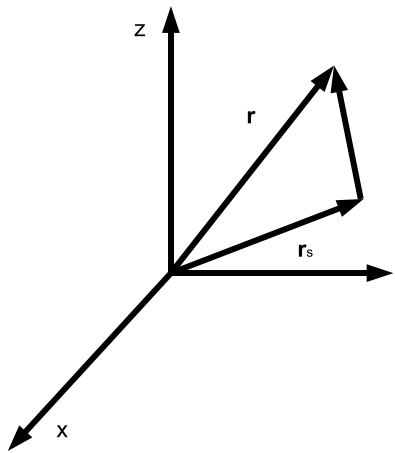}
\end{center}
\caption{Vector decomposition. }\label{fig2}
\end{figure}

\begin{equation}\label{eq1}
R^2 = [x - x_s(t - R/c)]^2 + [y - y_s(t - R/c)]^2 + [z - z_s(t - R/c)]^2.
\end{equation}

In the case of supersonic motion of the source, which moves along a straight line with constant speed, there are two emission points which will produce simultaneous contributions to the sound field at the observation point O. See Fig. 3.

\begin{figure}
\begin{center}
\includegraphics[width=0.4\textwidth]{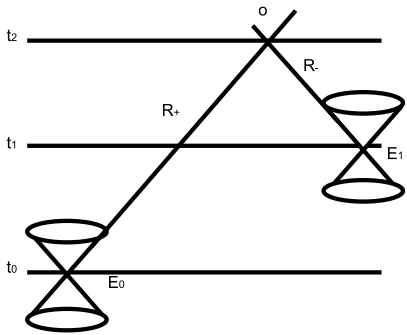}
\end{center}
\caption{Source in uniform supersonic motion. Here, $E_0$ and $E_1$ are the emission points and $R_+$ and $R_-$ are the two possible emission distances given by Eq. (\ref{eq2}) }\label{fig3}
\end{figure}

To simplify our calculations, let us assume that the source is moving along the $x$ axis with velocity $v$. Also, we assume that at $t = 0$ the source cross the origin of the coordinate system.  We get, $x_s(t) = vt$, $y_s = 0$ and $z_s = 0$. Eq. (\ref{eq1}) reads

\begin{equation}
R^2 = [x - v(t - R/c)]^2 + y^2 + z^2.
\end{equation}
This equation is satisfied by
\begin{equation}\label{eq2}
R = \frac{\frac{v}{c}(x - vt)\pm \sqrt{(x - vt)^2 + (1 - v^2/c^2)(y^2 + z^2)}}{(1 - \frac{v^2}{c^2})}.
\end{equation}
Note that, since the speed of the source is supersonic, both sign in Eq. (\ref{eq2}) are acceptable. There are one forward and one backward going wave reaching simultaneously the point of observation O, emitted at different times from $E_0$ and $E_1$, see \cite{rcmr86}.

\section{QUANTIZED FIELD INTERACTING WITH A CLASSICAL SUPERSONIC SOURCE}\label{sec3}

The equation of motion for the sound field is written in the form

\begin{equation}\label{eqfield}
\Box \psi(t, {\bf x}) = Q(t, {\bf x}),
\end{equation}
where $\psi$ is a quantum scalar field and the classical source can be thought as a small pulsating sphere  with source distribution density expressed by $Q(t, {\bf x}) = q(t)\delta(x - vt)\delta(y)\delta(z)$. This point source can be produced for instance by the heating and expansion  caused by some modulated radiation focused at a point that is moved through the fluid, or else by the interaction between neutral atoms and electrons, as mentioned in \cite{mi68}.

\begin{figure}
\begin{center}
\includegraphics[width=0.4\textwidth]{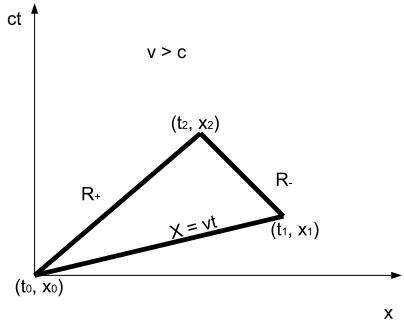}
\end{center}
\caption{Source in uniform supersonic motion. Here we show the event coordinates in spacetime. The Green function is 
evaluated at the point ($x_2, t_2$) due the the double emission that occurs at ($x_0, t_0$) and ($x_1, t_1$).}\label{fig4}
\end{figure}

We can write the field equation (\ref{eqfield}) in terms of a particular c-number

\begin{equation}\label{eqcnumber}
\psi(x) = \int d^4y G(x - y)Q(y)
\end{equation}
which is expressed in terms of a Green function $G(x - y)$ to the case illustrated in the Fig. (\ref{fig4}) when the source is supersonic. The Green function $G(x - y)$ satisfies
\begin{equation}\label{eqgreen}
\Box G(t_2, {\bf x}_2) = \delta(t_2 - t_0)\delta({\bf x}_2 - {\bf x}_0) + \delta(t_2 - t_1)\delta({\bf x}_2 - {\bf x}_1).
\end{equation}
The solution of Eq. (\ref{eqgreen}) can be given using a Fourier transform of the Green and Dirac delta functions.

Let us assume that we quantize the acoustic perturbations. For details see the Ref. \cite{landau}. The construction of the usual Fock space follows in the standard way for the phonons. Defining a unitary $S$ operator which links the in and out fields

\begin{equation}\label{equni1}
\psi_{out}(x) = S^{-1}\psi_{in}(x)S ,
\end{equation}
it will link the vacuum states $|0\rangle_{in}$ and $|0\rangle_{out}$ according to

\begin{equation}
|0\rangle_{out} = S^{-1}|0\rangle_{in} = S^{\dagger}|0\rangle_{in}
\end{equation}
or

\begin{equation}
|0\rangle_{in} = S|0\rangle_{out}
\end{equation}

In the case considered in this paper, the given source interacts with the quantized fluid  without reaction from it, that is, the system splits and ceases to be self-consistent. The $S$-matrix will be written in the normal order by

\begin{eqnarray}\label{eqs1}
S = e^{-i\int d^4x\psi_{in}(x)Q(x)} =  e^{-i\int d^4x\psi_{in}^{(-)}(x)Q(x)}e^{-i\int d^4x\psi_{in}^{(+)}(x)Q(x)}\\ \nonumber \times e^{-\frac{1}{2}\int\int [d^4xd^4y]\hspace{0.1mm}_{in}\langle 0|T\psi_{in}(x)Q(x)\psi_{in}(y)Q(y)|0\rangle_{in} }.
\end{eqnarray}
This form satisfies the unitarity condition claimed in Eq.(\ref{equni1}); we have also used the identity $e^Ae^B = e^{A+B+[A,B]/2}$ (which is valid whenever $[A,[A,B]] = [B,[A,B]] = 0$) as also a decomposition of $\psi_{in}(x) = \psi_{in}^{(+)}(x) + \psi_{in}^{(-)}(x)$, the annihilation and creation operators respectively.

In the last term in Eq. (\ref{eqs1}) we need the positive Wightman function $iG^{+}$, which can be evaluated from Eq. (\ref{eqgreen}). Thus,

\begin{eqnarray}
\hspace{0.1mm}_{in}\langle 0|T\psi_{in}(x)\psi_{in}(y)|0\rangle_{in}  = \int \frac{cd^4k}{(2\pi)^3}\theta(k^0)\delta(k^2) \\ \nonumber \times \left[e^{i{\bf k}.\Delta {\bf x'} - ick^0\Delta\tau'} + e^{i{\bf k}.\Delta {\bf x''} - ick^0\Delta\tau''}\right],
\end{eqnarray}
where we defined $\Delta {\bf x'} \equiv {\bf x}_2 - {\bf x}_0$, $\Delta\tau' \equiv t_2 - t_0$, $\Delta {\bf x''} \equiv {\bf x}_2 - {\bf x}_1$ and $\Delta\tau'' \equiv t_2 - t_1$.

The probability amplitude for the phonons to remain in the ground state is given by: $\hspace{0.1mm}_{out}\langle 0|0\rangle_{in} = \hspace{0.1mm}_{in}\langle 0|S|0\rangle_{in}$. To calculate the transition amplitude from the vacuum state to the corresponding state which emits $N$ phonons by the classical current $Q(x)$ we have the following: $\hspace{0.1mm}_{in}\langle N|S|0\rangle_{in}$. Where we have used the simplified notation $|N\rangle = |k_1\lambda_1,...,k_n\lambda_n\rangle $, and the probability $P(N)$ reads,

\begin{equation}\label{eqprob2}
P(N) = \sum_{\lambda=1,2}{\left|\hspace{0.1mm}_{in}\langle N_{{\bf k}\lambda}|S|0\rangle_{in}\right|^2},
\end{equation}
the sum being carried out over all sets of occupation numbers $N_{{\bf k}\lambda}$ for all values of momenta and polarizations.

It is straightforward to show that the last term in the exponential in the right hand side of Eq. (\ref{eqs1}) can be written as follows:

\begin{eqnarray}\label{eq13}
&\int\int [d^4xd^4y]\hspace{0.1mm}_{in}\langle 0|\psi_{in}(x)Q(x)\psi_{in}(y)Q(y)|0\rangle_{in} = -\frac{c}{2}\int \frac{d^4k}{(2\pi)^3}\delta(k^2) \times \\ \nonumber &\left\{\int d^4x_2Q(x_2)e^{-ik.x_2}\left[\int d^4x_0Q(x_0)e^{ik.x_0} + \int d^4x_1Q(x_1)e^{ik.x_1}\right]\right\},
\end{eqnarray}
where we have defined: $x_2 \equiv -{\bf x}_2 + t_2$, $x_0 \equiv -{\bf x}_0 + t_0$ and $x_1 \equiv -{\bf x}_1 + t_1$.

In Eq. (\ref{eq13}), it is possible to use the Fourier transform for the supersonic source $Q(x)$, which is 
$\tilde{Q}(k) = \int d^4xQ(x)e^{-ik.x}$:  where, only the sound-like arguments are nonzero, so that the matrix 
$S$ must be written in the form

\begin{equation}
S =  e^{-i\int d^4x\psi_{in}^{(-)}(x)Q(x)}e^{-i\int d^4x\psi_{in}^{(+)}(x)Q(x)}e^{-c\int{d\tilde{k}\left|\tilde{Q}(k)\right|^2_{k^0=ck}}}.
\end{equation}
with $d\tilde{k} = \frac{d^3k}{(2\pi)^32k} $.

Now, by using the Dyson formula given by the expression:$e^{-i\int_{-\infty}^{+\infty}dtH(t)} = \frac{1}{n!}\int_{-\infty}^{+\infty}dt_1...\int_{-\infty}^{+\infty}dt_nH(t_1)...H(t_n)$, it is possible to show that

\begin{equation}\label{eqbra}
\hspace{0.1mm}_{in}\langle N|S|0\rangle_{in} = e^{-\int d\tilde{k}|\tilde{Q}(k)|^2}\frac{(-i)^n}{n!}\int d^4x_1...\int d^4x_n\hspace{0.1mm}_{in}\langle N|\prod_{i=1}^nQ_i\psi_i^{(-)}(x)|0\rangle_{in}.
\end{equation}
This contribution is given by the creation operator.

On the other hand, only the annihilation operator does contribute to the other term in Eq. (\ref{eqprob2}):

\begin{equation}\label{eqket}
\hspace{0.1mm}_{in}\langle 0|S^*|N\rangle_{in} = e^{-\int d\tilde{k}|\tilde{Q}(k)|^2}\frac{(i)^n}{n!}\int d^4x_1...\int d^4x_n\hspace{0.1mm}_{in}\langle 0|\prod_{i=1}^nQ_i\psi_i^{(+)}(x)|N\rangle_{in}.
\end{equation}
Using Eq. (\ref{eqbra}) and Eq. (\ref{eqket}) in Eq. (\ref{eqprob2}), we get the following

\begin{eqnarray}\label{eq281}
P(N) = \frac{e^{-2\int d\tilde{k}|\tilde{Q}(k)|^2}}{(n!)^2}\int d^4x_1...d^4x_nd^4x_{n+1}...d^4x_{2n} \\ \nonumber \times \hspace{0.1mm}_{in}\langle 0|\prod_{i=1}^{n}Q_i(x)\psi^{(+)}_i(x)\prod_{i'=n+1}^{2n}Q_{i'}(x)\psi^{(-)}_{i'}(x)|0\rangle_{in}
\end{eqnarray}
where we have used that $\sum_{\lambda = 1,2}|N\rangle\langle N| = I$, quantity $I$ being the identity. So, Eq. (\ref{eq281}) can be written in the form

\begin{eqnarray}
P(N) = \frac{e^{-2\int d\tilde{k}|\tilde{Q}(k)|^2}}{(n!)^2}\int dx_1...dx_ndx_{n+1}...dx_{2n} \times \\ \nonumber \prod_{i=1}^N\prod_{i'=N+1}^{2N}Q_i(x)Q_{i'}(x)\hspace{0.1mm}_{in}\langle 0|\psi_i^{(+)}(x)\psi_{i'}^{(-)}(x)|0\rangle_{in}.
\end{eqnarray}

We get that $P(N)$ can be written as

\begin{eqnarray}
P(N) = \frac{e^{-2\int d\tilde{k}|\tilde{Q}(k)|^2}}{n!}\left[\int\int dxdyQ(x)Q(y)\hspace{0.1mm}_{in}\langle 0|\psi^{(+)}(x)\psi^{(-)}(y)|0\rangle_{in}\right]^n,
\end{eqnarray}
where the term $\hspace{0.1mm}_{in}\langle 0|\psi^{(+)}(x)\psi^{(-)}(y)|0\rangle_{in}$ is the positive Wightman function 
$G^+(x - y)$ already calculated. Following the same steps as before, we find $P(N)$ as

\begin{equation}\label{eqpoisson1}
P(N) = \frac{e^{-2\int d\tilde{k}|\tilde{Q}(k)|^2}}{n!}\left[2\int d\tilde{k}|\tilde{Q}(k)|^2\right]^n .
\end{equation}
On defining $\nu \equiv 2\int d\tilde{k}|\tilde{Q}(k)|^2$, we get

\begin{equation}\label{eqpoisson2}
P(N) = \frac{e^{-\nu}}{n!}(\nu)^n,
\end{equation}
which is a Poisson distribution.

Note that this result is very similar to the subsonic source's.  From this point of view, the problem is similar to the one of a classical sub-luminal source interacting with a quantized electromagnetic fields, which can be found in many books on QFT/QED (see for example \cite{qed} \cite{QED2}). The problem is equivalent to the one source case because only one phonon can be detected by the observer at a given time. So the Green equation reads

\begin{equation}
\Box G(x_2 - x_0) = \delta(t_2 - t_0)\delta^3({\bf x}_2 - {\bf x}_0).
\end{equation}

The probability distribution in this case, after performing the calculations as indicated above, is also a Poissonic distribution. The main difference is that the average number is defined as  $\nu \equiv \int d\tilde{k}|\tilde{Q}(k)|^2$, as it is already known in the electromagnetic case: Thus, it is possible to notice that $\nu_{supersonic} =  2\nu_{subsonic}$.

Up to now we did not discuss anything about the possibility of wave interference in the supersonic case. It is well known that if the sound waves amplitudes are summed, the interference is constructive; but if they are out of phase, this interference is partially destructive, and such an effect is expected to modify the result found in Eq. (\ref{eqpoisson1}).

If the sound strength is specified by $q = q_0e^{-i\omega t}$, using the result found in the chapter 11 of \cite{mi68}, we have that $\psi(t, {\bf x})$ can be written as

\begin{equation}
\psi(t, {\bf x}) = \frac{q_0e^{-i\omega t}}{4\pi R_1}\left[e^{i\omega R^+/c} + e^{i\omega R^-/c}\right] = \frac{q_0e^{-i\omega (t - R^+/c)}}{4\pi R_1}\left[1 + e^{i\frac{\omega}{c}( R^- - R^+)}\right].
\end{equation}
This is the classical solution of Eq. (\ref{eqfield}) with the sound strength given above.

\begin{figure}
\begin{center}
\includegraphics[width=0.4\textwidth]{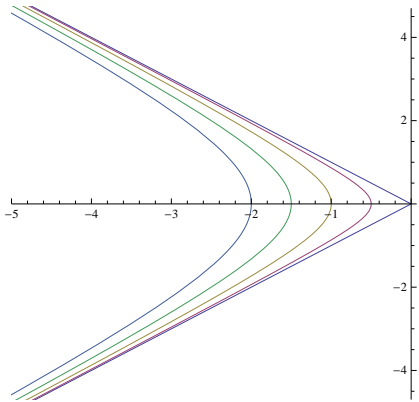}
\end{center}
\caption{The figure shows the interference pattern in 2 dimensions, given by the waves emitted by the supersonic source. 
Here we have five different values of $n = 0, 1,...,5$, with $v = 1.1c$. For the analogous superluminal case, see 
Refs. \cite{r74} and \cite{r98}}\label{fig5}
\end{figure}

Using Eq. (\ref{eq2}) we have 

\begin{equation}\label{eqhiper}
\psi(t, {\bf x}) = \frac{q_0e^{-i\omega (t - R^+/c)}}{4\pi R_1}\left[1 + e^{-2i\frac{\omega}{c}\frac{R_1}{M^2 - 1}}\right],
\end{equation}
where $R_1 = \sqrt{(vt - x)^2 - (M^2 - 1)(y^2 + z^2)}$ and $M = v/c$ is the Mach number. This, as expected\cite{r74, r98}, is the equation of a hyperboloid, with maximum and minimum values for the wave field given by $\frac{2\omega}{c}\frac{R_1}{M^2 - 1} = 2n\pi $ and $\frac{2\omega}{c}\frac{R_1}{M^2 - 1} = (2n + 1)\pi$, respectively. Such equation gives us Fig. (\ref{fig5}), which depicts the interference pattern measured by an observer at such instant of time ($t_2$). The waves were emitted by the supersonic source at the two different instants of time $t_0$ and $t_1$.

\section{SUMMARY AND DISCUSSIONS}\label{sec4}

In this paper we proposed an analog model for a tachyonic field in a fluid. The analogy consists in the study of a supersonic point-like source that emits phonons in a fluid at rest. The sound field inside the ``sound cone" can be regarded as a superposition of the two fields due to the two emissions points $E_{\pm}$: See Figs. (\ref{fig3}) and  (\ref{fig4}). By considering the interaction picture of the quantized acoustic perturbation with the external supersonic source, we found that the emission probability is a Poissonic distribution which shows the statistical independence of the emission of successive phonons. The main difference with the subsonic case refers to the average number of emitted phonons. In the supersonic case, we have shown that this number is twice as big as compared to the subsonic case ($\nu_{supersonic} = 2\nu_{subsonic}$).

A further interesting effect due the supersonic source is the interference of waves. It is well known that interference leads to regions of minimum and maximum amplitude in the fluid. In this paper this effect is illustrated in Fig. (\ref{fig5}). It is also possible to consider other effects of tachyon physics in fluids, as for example the behavior of a tachyonic field in curved space-time \cite{tss81}. 

\bigskip
\centerline{\bf ACKNOWLEDGMENTS}
The authors would like to thank E. Recami and S. Giovanazzi for helpful discussions. This work is partially supported 
by the Conselho Nacional de
Desenvolvimento Cient\'{\i}fico e Tecnol\'{o}gico - CNPq and
Coordena\c{c}\~ao de Aperfei\c{c}oamento de Pessoal de N\'{\i}vel
Superior - CAPES (Brazilian Research Agencies).

\end{document}